\documentclass[11pt,letterpaper]{article}
\usepackage{amsmath,bbm,amssymb,amsthm,graphics,epsfig,setspace,amsthm,hyperref,booktabs,float,adjustbox,amsthm}
\usepackage[flushleft]{threeparttable}
\usepackage{natbib}
\usepackage{xcolor}
\usepackage{authblk}
\usepackage[normalem]{ulem}
\usepackage{enumitem}
\usepackage{algorithmicx}
\usepackage{algorithm}
\usepackage{algpseudocode}
\usepackage{makecell}

\usepackage[margin=0.8in]{geometry}
\newtheorem{theorem}{Theorem}

\newtheorem{remark}{Remark}

\def\xx{{\bf x}}
\def\uu{{\bf u}}
\def\zz{{\bf z}}
\def\vv{{\bf v}}
\def\rr{{\bf r}}
\def\bbeta{{\boldsymbol\beta}}
\def\btheta{{\boldsymbol\theta}}

\newcommand{\re}{\mathbb{R}} 
\def\var{\text{Var}}

\DeclareMathOperator*{\argmin}{arg\,min}

\newcommand{\benu}{\begin{enumerate}}
	\newcommand{\eenu}{\end{enumerate}}
\newcommand{\bi}{\begin{itemize}}
	\newcommand{\ei}{\end{itemize}}

\usepackage{comment} 

\usepackage{graphicx}
\graphicspath{ {./images/} }

\begin{document}

\title{Censored broken adaptive ridge rank regression via induced smoothing}

\author[1]{Suyeon Seon}
\author[2]{Dipankar Bandyopadhyay}
\author[1,3]{Seongoh Park}
\author[1,3]{Dongha Kim}
\author[1,3]{Taehwa Choi\thanks{Corresponding author:  School of Mathematics, Statistics and Data Science, Sungshin Women’s University, 2, Bomun-ro 34da-gil, Seongbuk-gu, Seoul 02844, South Korea. E-mail: tchoi@sungshin.ac.kr}}

\date{} 

\affil[1]{\small School of Mathematics, Statistics and Data Science, Sungshin Women’s University, Seongbuk-gu, Seoul 02844, South Korea} 
\affil[2]{\small Department of Biostatistics, Virginia
Commonwealth University, Richmond, Virginia 23298-0032, USA } 
\affil[3]{\small Center for Data Science, Sungshin Women’s University, Seongbuk-gu, Seoul 02844, South Korea } 
 
\maketitle

\begin{abstract}

Broken adaptive ridge (BAR) penalty approximates $L_0$-regularization through iterative reweighting of $L_2$ penalties. This penalty enjoys both the oracle property and the grouping effect for highly correlated covariates, making it particularly attractive for penalized regression with complex dependence among predictors. In this paper, we develop a BAR-penalized linear rank regression method for the semiparametric accelerated failure time model with right-censored data. Computational tractability is achieved by applying induced smoothing to the nonsmooth Gehan-type rank estimating function, yielding a more stable framework for estimation and inference. For scalable penalization, we develop a cyclic coordinate descent algorithm that minimizes the penalized objective function, and estimates the regression coefficients in a coordinate-wise manner. We further extend the proposed method to more complex survival endpoints, such as multivariate partly interval-censored (PIC) data. Under mild conditions, the proposed estimator satisfies both the oracle property and the grouping effect, and the variance estimator of the informative coefficients can be derived in analytic form. Numerical studies using synthetic data compare our approach to several well-known penalties, and demonstrate its superior selection accuracy and estimation efficiency across various scenarios. Furthermore, applications to right-censored outcomes from primary biliary cirrhosis, and correlated PIC outcomes from colorectal cancer further illustrate the practical utility of the proposed method. The R package \texttt{aftPenCDA} for implementing the method is available on \texttt{R CRAN}. 


\end{abstract}

\section{Introduction}
\label{sec1}

Penalized regression has been widely adopted in many scientific applications to extract meaningful insights in the era of data deluge. As a result, extensive penalization methods have been developed, providing an effective framework to yield parsimonious and interpretable models by imposing a regularization term for selecting informative features. These approaches have proven essential in many fields where identifying relevant features from high-dimensional data is crucial for scientific discovery. Among them, the $L_0$-penalized regression, which penalizes the cardinality of the model, is a natural approach for variable selection \citep{breiman96}. However, its non-convex nature and lack of scalability makes it NP-hard and computationally infeasible. To address this problem, as a surrogate for $L_0$ penalization, $L_1$-based penalization approaches have received their attention, such as LASSO \citep{LASSO96}, and its numerous extensions, which includes smoothly clipped absolute deviation \citep[SCAD;][]{SCAD01}, adaptive LASSO \citep[ALASSO;][]{Adapt06}, and minimax concave penalty \citep[MCP;][]{MCP10}.

Recently, the broken adaptive ridge (BAR) penalization, which approximates the $L_0$ penalty via reweighted $L_2$ penalization \citep{frommlet16}, has been rigorously studied from both computational and theoretical perspectives \citep{Dai2018}. The iterative reweighting procedure approximates the behavior of the $L_0$-penalization as the iterations proceed. One of the most intriguing theoretical aspects of the BAR penalization is that the resulting estimator satisfies not only the oracle property but also the grouping effect, whereby, the estimated coefficients of highly correlated features tend to have similar magnitudes. Furthermore, the BAR procedure often yields a sparser model than other penalization methods without compromising prediction performance. These attractive features have led to various extensions of the BAR framework, particularly in generalized linear models \citep{li2021} and survival models, with most survival-related developments centered on classical Cox-based formulations. \cite{Eric2020} proposed a Cox-based BAR procedure specifically designed for right-censored data with massive sample sizes, as well as extending this approach \cite{kawa21} to accommodate large-scale competing risks data. Furthermore, \cite{zhao20} utilized a sieve maximum likelihood estimation for the interval-censored (IC) data.

As an alternative to the Cox model, the semiparametric accelerated failure time (AFT) model has become increasingly attractive because it models covariate effects directly on the logarithm of survival time while leaving the error distribution unspecified \citep{BJ1979,ts90,jin03,JinLin2006}. This formulation offers both structural simplicity and an intuitive interpretation in terms of event-time acceleration or deceleration. Building on this framework, various $L_1$-based penalization methods have been proposed over the past several decades, including the Buckley-James framework \citep{wang08,johnson08}, rank-based approaches \citep{xu2010,Choi21} and censored quantile regression \citep{son22}. For the BAR penalization, \cite{sun22} applied the BAR penalty to the semiparametric AFT model, in which the censored observations are imputed with the Leurgans' response transformation \citep{luer87}. Later, \cite{Leechoi24} adopted the Buckley-James framework, which imputes the censored observations by their conditional expectation, and extended it to double censoring. Although these methods adopt least-squares-based computation, they require iterative updating between the regression parameters and the distribution function due to the inherent semiparametric assumption. Furthermore, inferential procedures for the regression parameters, which constitute a crucial aspect of survival analysis, have not been rigorously discussed.

In this paper, we propose a novel BAR procedure for the semiparametric AFT model by utilizing the rank-based estimating equation \citep{ts90}. A key advantage of the rank-based approach is its ability to circumvent nuisance parameter estimation, as inference is driven solely by the residual ranks rather than by error density specification. Consequently, the method yields a computationally elegant estimating procedure, often formulated through an $L_1$-type objective function \citep{jin03}. Unlike the $L_1$-type penalized rank regression approaches \citep{xu2010,Choi21}, incorporating an $L_2$-type penalty into the $L_1$-type objective function substantially increases the computational burden. As a remedy, we adopt the induced smoothing \citep[IS;][]{Brown07} approach to smooth the estimating function, and ensure its differentiability. There have been numerous studies on the smoothed analysis of rank-based regression \citep{Heller07,Choi21}, which, however, require a carefully chosen bandwidth. In contrast, the IS method does not require an explicit bandwidth selection, as the bandwidth is implicitly and optimally determined by the data.  

To enhance the computational efficiency, we employ a cyclic coordinate descent algorithm \citep{wright2015coordinate} that sequentially updates the coefficients along each coordinate direction. To our knowledge, penalization within the IS framework has not yet been explored. Hence, building on the proposed algorithm, we provide a unified framework that can readily accommodate existing penalties. Desired properties of the BAR penalty, including the oracle property and the grouping effect, are rigorously justified. Our method provides simple variance estimation procedure for the informative coefficients with an analytic formula. We further extend the method to the clustered/correlated partly IC (PIC) data, also frequently encountered in biomedical settings. Under this setting, some event times are observed exactly, while others are IC, meaning that the true failure time is unobserved but known to lie within a prespecified examination interval, in addition to accommodating left- and right-censored observations as additional censoring patterns. To incorporate possible informative cluster size \citep[ICS;][]{anyaso2023pseudo} scenario -- often a salient feature in correlated data settings where the number of observations within a cluster may be associated with the survival outcomes of the clustered units, we adopt the marginal analysis approach of the multivariate AFT model adjusting for the cluster weights in the objective function \citep{wang08,choi2025rank}. 

The rest of this paper is organized as follows. In Section \ref{sec2}, the induced (kernel) smoothing broken adaptive ridge (IS-BAR) procedure is proposed, along with an efficient coordinate descent algorithm. Theoretical results, including the oracle property and the grouping effect, are presented, along with an efficient resampling-based variance estimation procedure. Furthermore, we extend the method to the multivariate PIC data by incorporating ICS. Section \ref{sec3} demonstrates the finite sample performances of the proposed method via Monte-Carlo simulation studies, while Section \ref{sec4} illustrates the proposed methodology via application to two real-life data examples.
Final remarks and discussions are made in Section \ref{sec5}. All technical details are postponed to the Supplementary materials.  The \texttt{R} package \texttt{aftPenCDA} implementing our method is available on CRAN.

\section{Statistical Model} \label{sec2}

\subsection{Induced smoothing for logrank regression} \label{ss:insmooth}

Suppose that we observe a random sample of $n$ subjects. For $i=1,\ldots,n$, let $T_i$ and $C_i$ be the failure time and the censoring time, respectively, and $\xx_i$ be $p$-vector of covariates for the $i$th subject. Then, the semiparametric AFT model is given by
\begin{equation}\label{eq:aft} 
    \log T_i =  \xx_i^T\bbeta + \varepsilon_i,~~ (i=1,\ldots, n),
\end{equation}
where, $\bbeta$ is a $p$-dimensional vector of unknown regression parameters and $\varepsilon_i$ is an independently and identically distributed random error from an unspecified distribution function. When the data are subject to right-censoring, the observed sample is represented as $(\tilde T_i, \delta_i, \xx_i),~i=1,\ldots,n$,
where $\tilde T_i = \min(T_i, C_i)$ is the observed failure time, and $\delta_i=I(T_i<C_i)$ is the censoring indicator with the indicator function $I(\cdot)$. We assume that the censoring time $C_i$ is conditionally independent of the failure time $T_i$ given $\xx_i$.

Following \cite{ts90}, the weighted log-rank estimating equation estimates the regression parameter $\bbeta$,
$$
\frac1n \sum_{i=1}^n\psi_i(\bbeta)  \delta_i 
\left[
\xx_i - \frac{\sum_{j=1}^n \xx_j I\{ e_j(\bbeta) > e_i(\bbeta) \} }{\sum_{j=1}^n I\{ e_j(\bbeta) >e_i(\bbeta) \} }
\right],
$$
where, $e_i(\bbeta) = \log \tilde T_i - \xx_i^T\bbeta$ and $\psi_i(\cdot)$ is a non-negative possibly data-dependent weight function. Unfortunately, this estimating equation is not only discontinuous but also non-monotone with respect to $\bbeta$, leading to possibly multiple roots to the equation. Adopting the weight $\psi_i(\bbeta) = n^{-1}\sum_{j=1}^n I\{ e_j(\bbeta) > e_i(\bbeta) \}$ \citep{Fygen94}, the above estimating equation becomes a monotone function, with 
\begin{align}\label{eq:useq}
\tilde S_n(\bbeta) = \frac1{n^2} \sum_{i=1}^n\sum_{j=1}^n \delta_{i} (\xx_i-\xx_j) I\{e_i(\bbeta) < e_j(\bbeta)\},
\end{align}
serving as a gradient of the following convex objective function
$$
\tilde L_n (\bbeta) = \frac1{n^2} \sum_{i=1}^n\sum_{j=1}^n \delta_i \{ e_i(\bbeta)-e_j(\bbeta) \}_-,
$$
where, $\{u\}_- = |u|I(u<0)$. Optimizing the objective function $\tilde L_n (\bbeta)$ is straightforward using linear programming \citep{jin03}.
Unfortunately, since the objective function $\tilde L_n (\bbeta)$ is still discontinuous in $\bbeta$, the computation time grows exponentially as the sample size increases.
As an alternative, we utilize the IS estimating equation that is asymptotically equivalent to the unsmoothed estimating equation \eqref{eq:useq} \citep{Brown07, JohnL09}.

Specifically, let $\zz\sim N({\bf0},I_p)$ be a random variable independent of the data, where $I_p$ denotes the $p$-dimensional identity matrix.
Let $\Sigma$ be a $p$-dimensional symmetric and positive definite matrix satisfying $\|\Sigma\|=O(n^{-1})$. The smoothed estimating equation can be constructed by taking the expectation of the unsmoothed estimating equation \eqref{eq:useq} after adding the perturbation term $\Sigma^{1/2} \zz$ to the regression parameter $\bbeta$, i.e., $ S_n(\bbeta) = E_\zz\{\tilde S_n(\bbeta+ \Sigma^{1/2} \zz)\}$. With some simple algebra, the smoothed estimating equation can be written as
\begin{align}\label{eq:iseq}
 S_n(\bbeta) = \frac1{n^2} \sum_{i=1}^n\sum_{j=1}^n\delta_i (\xx_i-\xx_j) 
	\Phi \bigg( \frac{e_i(\bbeta)-e_j(\bbeta)}{r_{ij}} \bigg),
\end{align}
where, $\Phi(\cdot)$ is the standard normal cumulative distribution function (cdf) and $r_{ij}^2 = (\xx_i-\xx_j)^T \Sigma (\xx_i-\xx_j)$.  Furthermore, by the property of the standard normal cdf,  the smoothed estimating equation \eqref{eq:iseq} is the gradient of the following smoothed objective function,
\begin{align} \label{eq:isobj}
L_n(\bbeta) = \frac1{n^2} \sum_{i=1}^n\sum_{j=1}^n \delta_i \bigg[
	\{e_j(\bbeta) - e_i(\bbeta) \} \Phi \bigg( \frac{e_i(\bbeta)-e_j(\bbeta)}{r_{ij}} \bigg)
	+ r_{ij} \phi \bigg( \frac{e_i(\bbeta)-e_j(\bbeta)}{r_{ij}} \bigg)\bigg],
\end{align}
and has the negative derivative as
\begin{equation}\label{eq:Amat}
    A_n (\bbeta) =\frac{1}{n^2}\sum_{i=1}^{n}\sum_{j=1}^{n}\delta_i\phi\left(\frac{e_i(\bbeta) - e_j(\bbeta)}{r_{ij}}\right)\frac{(\xx_i-\xx_j)(\xx_i-\xx_j)^{T}}{r_{ij}} ,
\end{equation}
where, $\phi(\cdot)$ is the density function of the standard normal distribution. Since \eqref{eq:isobj} is convex and continuously differentiable with respect to $\bbeta$, the root of the score function \eqref{eq:iseq}, denoted by $\tilde\bbeta$, exists and is unique.
Note that the asymptotic results of the IC regression estimator were rigorously proved \citep{Brown07,JohnL09}.
Specifically, $\Sigma$ can be consistently estimated by $\Sigma_n(\tilde\bbeta) = n^{-1}  A_n(\tilde\bbeta)^{-1} \Gamma_n(\tilde\bbeta) A_n(\tilde\bbeta)^{-1}$,
where $\Gamma_{n}(\bbeta) = n^{-1}\sum_{i=1}^n \xi_i(\bbeta)^{\otimes2}$,
${\bf a}^{\otimes2 }={\bf a}{\bf a}^T$
and 
\begin{align} \label{eq:xi}
\begin{split}
\xi_i(\bbeta) = \frac1{n}\sum_{j=1}^n \bigg[ &\delta_i (\xx_i-\xx_j) I\{e_i(\bbeta)< e_j(\bbeta) \}  \\
& ~~~~-\delta_j \frac{\sum_{r=1}^n (\xx_i-\xx_r)I\{e_r(\bbeta)>e_j(\bbeta) \} }{\sum_{m=1}^n I\{e_m(\bbeta)>e_j(\bbeta) \}} 
I\{e_i(\bbeta)\ge e_j(\bbeta) \}\bigg].    
\end{split}
\end{align}

\begin{remark}
The exact form \eqref{eq:xi} may be computationally inefficient as the sample size increases. Hence, a computationally efficient resampling approach \citep{resamp08} can be a reliable alternative.  To implement the resampling approach for calculating $\Gamma_n(\bbeta)$, we first generate a perturbation random sample $(R_{b1},\ldots, R_{bn})$, $b=1,\dots,B$, where $R_{bi} \sim \mathrm{Exp}(1)$. Next, the perturbed estimating function can be calculated by
$$
n^{1/2}S_n(\bbeta;R_b) = n^{-3/2}\sum_{i=1}^n\sum_{j=1}^nR_{bi}\delta_i(\xx_i -\xx_j)I\{e_i(\bbeta) < e_j(\bbeta)\}, ~~b = 1,\dots, B.
$$
We then obtain $\Gamma_n(\tilde\bbeta) = \var(n^{1/2}S_n(\tilde\bbeta;R_b))$ to approximate $\Gamma$. In our experience, this approach is faster than the direct calculation of \eqref{eq:xi} when the sample size is large, and this approach can also calculate the variance estimator of $\tilde \bbeta$ reliably.
\end{remark}

\subsection{Broken adaptive ridge estimation via induced smoothing} \label{ss:BAR}

To regularize the induced kernel smoothing broken adaptive ridge (IS-BAR) estimator, we consider the following minimization problem
\begin{equation}\label{eq:barobj}
    \eta(\bbeta^o)=\argmin_{\bbeta\in\mathbb{B}} \left\{ L_n(\bbeta) + \frac\lambda2 \sum_{j=1}^p {\beta_j^2}/{\beta_j^{o2}} \right\},
\end{equation}
where, the parameter space $\mathbb{B}$ is a known compact set in $\re^p$. 
For $k\ge 1$, let $\hat\bbeta^{(k)} = \eta(\hat\bbeta^{(k-1)})$,
where $\hat\bbeta^{(0)}$ is the unpenalized IC estimator $\tilde\bbeta$ -- the solution of \eqref{eq:iseq} set to zero. Then, the IS-BAR estimator is $\hat\bbeta = \lim_{k\to\infty}\hat\bbeta^{(k)}$, which can be obtained via standard Newton-type algorithms, which, however, involve numerically inefficient iterative estimating procedures between $\bbeta$ and $\Sigma$.

Motivated by the second-order approximation \citep{hao07,Choi21}, we consider an efficient cyclic coordinate descent algorithm, which successively minimizes the penalized objective function along the coordinate directions to reach the minima.
Using the unpenalized IC estimator $\tilde\bbeta$, we first calculate $S_n(\tilde\bbeta)$ and $A_n(\tilde\bbeta)$.
Then, by utilizing the Cholesky decomposition of $A_n(\tilde\bbeta) = \vv^T \vv$, the pseudo-response $\uu = (\vv^T)^{-1}\{A_n(\tilde\bbeta)\tilde\bbeta - S_n(\tilde\bbeta)\}$ and the pseudo-covariates $\vv$ are generated. Denote the standardized version of $\vv$ by $\vv_{*}$ such that $\vv_{*}^{T}\vv_{*} = n$, and the centralized version of $\uu$ by $\uu_{*}$. Then, the minimizer of the objective function \eqref{eq:isobj} is equivalent to the least-squares estimator by regressing $\uu$ on $\vv$, such that $\tilde\bbeta=n^{-1}\vv_*^T\uu_*$, i.e., the BAR-penalized objective function \eqref{eq:barobj} can be equivalently expressed as
$$
\eta(\bbeta^o) = \argmin_{\bbeta\in\mathbb B} \left\{\frac1{2} (\uu_*-\vv_*^T\bbeta)^T(\uu_*-\vv_*^T\bbeta) + \frac\lambda2 \sum_{j=1}^p \beta_j^2/\beta_j^{o2}
\right\}.
$$
It is worth noting that some of the $\beta_j$'s converge to zero in their limits,
and thus the adaptive weights $1/\beta_j^{o2}$'s may result in arithmetic overflow due to the denominator becoming zero \citep{Dai2018,kawa21}. Hence, we add a small perturbation to the weights to avoid numerical instability, i.e., $\lambda/2 \sum_{j=1}^p \beta_j^2/(\beta_j^{o2} + \gamma^2)$. We set $\gamma = 10^{-5}$.

To accommodate efficient computation, we utilize a cyclic coordinate descent algorithm that updates each coordinate of the penalized objective function. The efficiency of this approach has been confirmed by previous studies \citep{kawa21,Leechoi24} by updating the component-wise coefficient vector, $\hat\beta_j^{(k)}$, in an analytic-form expression. To this end, we use the following formulation for the coordinate-wise update (see derivation in the Supplementary Materials, Section 2),
$$
\hat{\beta}_j^{(k)} = \left\{\alpha_j^{(k)}  /2 + \sqrt{(\alpha_j^{(k)})^2 /4 - \lambda/n}\right\} I(|\alpha_j^{(k)}| > 2\sqrt{\lambda/n} ),
$$
where the sparsity of the coefficient is determined by the magnitude of $\alpha_j^{(k)}=n^{-1}\vv_{*,j}^T (\uu_* - \vv_*^T\hat\bbeta^{(k-1)}) + \hat\beta_j^{(k-1)}$. We postpone the technical details of the updating formula to the web-appendix.
Here, $\vv_{*,j}$ represents the $j$th column vector of $\vv_*$, i.e., $\vv_* = (\vv_{*,1}, \ldots,\vv_{*,p})$.
Our algorithm requires only a single matrix computation to obtain the pseudo-response and pseudo-covariates, whereas optimizing the original objective function \eqref{eq:barobj} entails multiple matrix computations, which becomes inefficient as the data dimension increases. 
We further remark that the proposed algorithm can be easily extended to incorporate other common penalties, such as LASSO and SCAD; this generalization is deferred to the Supplementary Materials (Section 3).

The proposed cyclic coordinate descent algorithm for the IS-BAR estimator is summarized in Algorithm 1. This algorithm proceeds with the iteration until either the maximum number of iterations ($k_{\rm max}= 100$) is reached, or the convergence criterion $\max_{j=1,\ldots,p} |\hat\beta_j^{(k)} - \hat\beta_j^{(k-1)}| < \epsilon=10^{-8}$ is met.
After the algorithm terminates, the final regression estimators are transformed back to the original scale. Empirically, the algorithm exhibits strong numerical stability and converges substantially faster than direct optimization \eqref{eq:barobj}.

\begin{algorithm}[H]
\caption{Coordinate descent algorithm for the IS-BAR}
\small
\begin{algorithmic}[1]
  \Require $\{(\tilde{T}_i,\delta_i, \xx_i)\},~i=1,\dots,n$: data, $\lambda$: tuning parameter
  \State Initialize $\hat{\boldsymbol{\beta}}^{(0)}$ based on IC estimators.
  \State Calculate the centralized pseudo-response $\uu_*$ and the standardized pseudo-covariates $\vv_*$ by $\hat\bbeta^{(0)}$.
  \State Initialize the residual: $\boldsymbol{\rr}^{(0)} = \boldsymbol{\mathrm{u}}_* - \boldsymbol{\mathrm{\mathrm{v}}}_*\hat{\boldsymbol{\beta}}^{(0)}$
  \For{$k=1,2,\ldots,$ $k_{\text{max}}=100$}
    \For{$j=1,2,\ldots,p$}
      \State Compute the partial residual: $\alpha_j^{(k)} 
        = n^{-1}\boldsymbol{\mathrm{v}}_{*,j}^{T} \boldsymbol{\rr}^{(k-1)} + \hat{\beta}_j^{(k-1)}$
      \State Update the IS-BAR estimator: $$\hat{\beta}_j^{(k)} = \left\{\alpha_j^{(k)}  /2 + \sqrt{(\alpha_j^{(k)})^2 /4 - \lambda/n}\right\} I(|\alpha_j^{(k)}| > 2\sqrt{\lambda/n} )$$
      \State Update the residual: $
        \boldsymbol{\rr}^{(k)} = \boldsymbol{\rr}^{(k-1)} - \vv_{*,j}\bigl(\hat{\beta}_j^{(k)} - \hat{\beta}_j^{(k-1)}\bigr)$
    \EndFor
    \If{$\max_{j=1,\ldots,p} |\hat\beta_j^{(k)} - \hat\beta_j^{(k-1)}| < 10^{-8}$}
    \State Break
   \EndIf
  \EndFor
 \State $\hat{\boldsymbol{\beta}} = \hat{\boldsymbol{\beta}}^{(k)}$
\end{algorithmic}
\end{algorithm}

\begin{remark}
For the optimal shrinkage parameter selection $\lambda$, we adopt the Bayesian information criterion (BIC). Denote $\hat{\bbeta}_\lambda$ as the IS-BAR estimator with shrinkage parameter $\lambda$. Similar to \citep{xu2010}, we derive the modified BIC as
$$\mathrm{BIC}_\lambda = T_\lambda + \log n  \cdot df_\lambda,$$
where, $T_\lambda = S_n(\hat{\bbeta}_\lambda)\Gamma^{-1}_n(\hat{\bbeta}_\lambda)S_n(\hat{\bbeta}_\lambda)$ and $df_\lambda$ is the number of non-zero components in $\hat{\bbeta}_\lambda$. The optimal tuning parameter $\lambda$ is determined by minimizing the $\mathrm{BIC}_\lambda$.
Note that each component of $T_\lambda$ converges to the unsmoothed quantities, as in \cite{xu2010}, and the optimal $\lambda$, minimizing the BIC, guarantees the correspondence to the true model.
\end{remark}

\subsection{Asymptotic results} \label{ss:theory}

To establish the oracle properties and the grouping effect of the proposed IS-BAR estimator, we begin by imposing the following regularity conditions.

\begin{description}
\item[\rm (A1)] The covariate $\xx$ is uniformly bounded almost surely, and the true parameter $\bbeta_0$ lies in a compact set $\mathbb{B}$. 
\item[\rm (A2)] The distribution function $F_0(\cdot)$ is uniformly bounded away from 0 and has  a density $f_0(\cdot)$ with continuous derivative bounded away from 0 on their support.
\item[\rm (A3)] The marginal distribution of $C$ is absolutely continuous, and has a bounded density $g_0(\cdot)$ on $\re$.
\item[\rm (A4)] The asymptotic slope matrix $A$ is nonsingular. In addition, there exist a constant $C_1$ such that 
$0 < 1/C_1 < \zeta_{min}\{A_n(\bbeta)\} \le \zeta_{max}\{A_n(\bbeta)\}< C_1 < \infty$
for some constant $C_1$, where $\zeta_{min}(\cdot)$ and $\zeta_{max}(\cdot)$ represents the smallest and the largest eigenvalues of a matrix, respectively.
\item[\rm (A5)] As $n$ goes to infinity, 
the tuning parameter $\lambda  \to \infty$ and $n^{-1/2}\lambda  \to 0$.
\end{description}
Conditions (A1)--(A4) are the regularity conditions from \cite{JohnL09}, which are classical assumptions in semiparametric AFT models. According to their work, $\tilde\bbeta$ is defined as a solution to $S_n(\bbeta) =0$ and is a strongly consistent estimator for $\bbeta_0$.
In addition, the term $n^{1/2} (\tilde\bbeta-\bbeta_0)$ converges to the mean-zero normal distribution with finite variance $A^{-1}\Gamma A^{-1}$, where the exact forms of $A$ and $\Gamma$ are given in the web-appendix. Conditions (A4) and (A5) are required to demonstrate the asymptotic results of the IS-BAR estimator. Under these conditions, the IS-BAR estimator satisfies the oracle property and the grouping effect \citep{Dai2018,li2021,Leechoi24}.

Before declaring the theorems, we first introduce some notations.  Let $\bbeta_0 = (\bbeta_{10}^T,\bbeta_{20}^T)^T$ be the true regression coefficient, 
where $\bbeta_{10}$ is the first $q$-vector of non-zero coefficients and $\bbeta_{20}$ is the remaining $(p-q)$-vector of zero coefficients.
Likewise, denote $\hat\bbeta^{(k)}=g(\bbeta^{(k)}) =  (g_1(\bbeta^{(k)})^T,g_2(\bbeta^{(k)})^T)^T$ and $\hat\bbeta = (\hat\bbeta_1^T,\hat\bbeta_2^T)^T$ be the decomposition of $\hat\bbeta^{(k)}$ and $\hat\bbeta$, respectively.  Define $L_{n}(\btheta;\lambda) = L_{n}(\btheta) + \frac{\lambda }{2}\btheta^T D(\bbeta)\btheta$, where $\btheta$ is a $p$-vector. Given $\bbeta_{20}=0$, define the penalized objective function of the reduced model as $L_{n1}(\btheta_1;\lambda) = L_{n1}(\btheta_1) + \frac{\lambda }{2}\btheta_1^T D_1(\bbeta_1)\btheta_1$, where $\btheta_1$ is a $q$-vector and $D_1(\bbeta_1)$ is the first $q\times q$-diagonal matrix of  $D(\bbeta) = \mathrm{diag}(\beta_1^{-2},\beta_2^{-2},\dots,\beta_{p}^{-2})$. The first derivative of $L_{n1}(\btheta_1;\lambda)$ is $L_{n1}'(\btheta_1;\lambda) = S_{n1}(\btheta_1;\lambda)= S_{n1}(\btheta_1) + \lambda D_1(\bbeta_1)\btheta_1$.

\begin{theorem} \label{thm1}
  Under the regularity conditions (A1)-(A5), with probability tending to 1 as $n \to \infty$, the IS-BAR estimator $\hat{\bbeta} = (\hat{\bbeta}_1^T, \hat{\bbeta}_2^T)^T$ satisfies the following properties:
  \begin{enumerate}[label=(\alph*)]
  	\item $\hat{\bbeta} = (\hat{\bbeta}_1^T, \hat{\bbeta}_2^T)^T$ exists and is unique with $\hat{\bbeta}_2 =0 $ and $\hat{\bbeta}_1$ being the unique fixed point of $f(\bbeta_1)$, which is a solution to $S_{n1}(\boldsymbol{\theta}_1;\lambda)=0$;
  	\item $n^{1/2}(\hat{\bbeta}_1 - \bbeta_{10}) \to_d N(0,A_1^{-1}\Gamma_1A_1^{-1}).$
  \end{enumerate}   
\end{theorem}

Theorem \ref{thm1} states the oracle property of the IS-BAR estimator. Specifically, Theorem \ref{thm1}-(a) establishes selection consistency of the IS-BAR estimator, and Theorem \ref{thm1}-(b) shows that the non-zero components of the BAR estimator follow the normal distribution,  asymptotically. This theorem can be proved by the limiting properties of the IC estimator and the Cauchy-Schwarz inequality. For the variance estimation, we split the IS-BAR estimator into two parts: $\hat\bbeta = (\hat\bbeta_1^T , \hat\bbeta_2^T)^T$, where $\hat\bbeta_1$ is a $q$-vector of informative coefficients and $\hat\bbeta_2$ is a $(p-q)$-vector of non-informative coefficients. In addition, let $A_{n1}(\hat\bbeta_1)$ and $\Gamma_{n1}(\hat\bbeta_1)$ be the $q\times q$ sub-matrices of $A_{n}(\hat\bbeta)$ and $\Gamma_{n}(\hat\bbeta)$, respectively.
Then, the variance estimator for $\hat\bbeta_1$ can be obtained as
$n^{-1} \Sigma_{n1}(\hat\bbeta_1) = n^{-1}A_{n1}(\hat\bbeta_1)^{-1}
\Gamma_{n1}(\hat\bbeta_1)A_{n1}(\hat\bbeta_1)^{-1}$.

\begin{theorem}\label{thm2}
Assume that the columns $\xx_r, r=1,\ldots,p$ of matrix $\boldsymbol{\mathrm{\xx}}$ are standardized, such that $\sum_{i=1}^n x_{ir} = 0$ and $\xx^T_r\xx_r =n$. Let $\hat{\bbeta}$ be the IS-BAR estimator. Under regularity conditions (A1)-(A5), for any $r < s$ such that $\hat{\bbeta}_r\hat{\bbeta}_s \ne 0$, and with probability tending to 1, 
$$|\hat{\bbeta}_r^{-1} - \hat{\bbeta}_s^{-1}| \le  \frac1\lambda \sqrt{4(1- \rho_{rs}) e_n},$$
where, $e_n = \sum_{i=1}^n\delta_i$ and $\rho_{rs} = \frac1n\xx_r^T\xx_s$ is the sample correlation of $\xx_r$ and $\xx_s$.
\end{theorem}
Theorem \ref{thm2} establishes the grouping effect of the IS-BAR estimator, which implies that the estimated coefficients of two highly correlated variables will be similar in magnitude. The proof of this theorem can be completed by bounding the difference between the coordinate-wise penalized estimating functions and the sample correlation of the covariates. Detailed proofs of the asymptotic results are presented in the Supplementary Materials (Section 1).

\subsection{Extension to the clustered partly interval-censored data}
\label{sec3}

In this section, we extend our method to accommodate clustered/correlated PIC data. This data structure can be viewed as a mixture of IC and exact observations arising from multiple clusters (e.g., cohorts), a form commonly encountered in large-scale clinical studies. Suppose there are $n$ clusters in the cohorts, and the $i$th cluster consists of $m_i$ patients. Then, model \eqref{eq:aft} can be extended to the marginal AFT model
\begin{align}\label{eq:maft} 
    \log T_{ik} = \xx_{ik}^T\bbeta + \varepsilon_{ik},~~(i=1,\ldots,n;k=1,\ldots,m_i),
\end{align}
where, $(\varepsilon_{i1},\ldots,\varepsilon_{im_i})$ are independent random vectors for $i=1,\ldots,n$. Within the $i$th cluster, the error terms, $\varepsilon_{i1},\dots,\varepsilon_{im_i},$ are assumed to be exchangeable with a common marginal distribution $F$. Note that model \eqref{eq:maft} reduces to the model \eqref{eq:aft} when $m_i = 1$ for all $i$. 

Under the PIC situation, some observations are subject to IC. Denote $\Delta_{ik}$ as an interval-censoring indicator, representing $\Delta_{ik}=1$ for an exact observation and $\Delta_{ik}=0$ for an IC observation.
Let $U_{ik}$ and $V_{ik}$ be the lower and upper bounds of the observation interval, respectively, such that $U_{ik} < T_{ik} < V_{ik}$.  $U_{ik}$ and $V_{ik}$ are observed when $\Delta_{ik}=0$. We further denote the observed bounds as $\tilde U_{ik} = \Delta_{ik}T_{ik} + (1-\Delta_{ik})U_{ik}$ and $\tilde V_{ik} = \Delta_{ik}T_{ik} + (1-\Delta_{ik})V_{ik}$. 
Then, the observed tuple consists of $(\tilde{U}_{ik}, \tilde{V}_{ik}, \Delta_{ik}, \xx_{ik})$ for $i=1,\ldots,n; k=1,\ldots,m_i$. In this setup, we assume that $U_{ik}$ and $V_{ik}$ are independent of $T_{ik}$ given $\xx_{ik}$, which implies that the censoring times do not provide additional information about the distribution of $T_{ik}$ given $\xx_{ik}$, thereby ensuring non-informative censoring. We further assume that the proportion of exact times is non-negligible, i.e., $P(\Delta_{ik}=  1) > 0$ for all $i$ and $k$. We can easily extend the principle of the right-censored rank regression to the partly IC situation by investigating the rank comparable pairs. Specifically, when the finite observed upper bound of a $k$th object in the $i$th cluster is smaller than the positive lower bound of a $l$th object in the $j$th cluster, i.e., $\{\Delta_{ik} + (1-\Delta_{ik})I(V_{ik}<\infty)\} \tilde V_{ik} \le \{\Delta_{jl} + (1-\Delta_{jl})I(U_{jl}>0)\} \tilde U_{jl}$, then this implies the same rank between two latent failure times $T_{ik} < T_{jl}$.
Let $u_{ik}(\bbeta) = \log\tilde{U}_{ik} - \xx_{ik}^T\bbeta$ and $v_{ik}(\bbeta) = \log\tilde{V}_{ik} - \xx_{ik}^T\bbeta$ denote the observed residuals under the model \eqref{eq:maft}. Then, we can extend the IS estimating equation in \eqref{eq:iseq} to the multivariate PIC data as
\begin{align}\label{eq:mpic_is}
    S_n^{\dagger}(\bbeta) = \frac1{n^2}\sum_{i=1}^{n}\sum_{k=1}^{m_i}\sum_{j=1}^{n}\sum_{l=1}^{m_j}\varphi_i\varphi_{j}\eta_{2ik}\eta_{1jl}(\xx_{ik} - \xx_{jl})\Phi\left(\frac{v_{ik}(\bbeta) - u_{jl}(\bbeta)}{r_{ikjl}}\right)
\end{align}
where, $r_{ikjl} = (\xx_{ik} - \xx_{jl})^T\Sigma(\xx_{ik} - \xx_{jl}), \eta_{2ik} = \Delta_{ik}+(1-\Delta_{ik})I(V_{ik} < \infty)$ and $\eta_{1jl} = \Delta_{jl} + (1-\Delta_{jl})I(U_{jl} > 0)$.
Note, when $m_i=1$ for all $i$, the equation \eqref{eq:mpic_is} reduces to the equation \eqref{eq:iseq}. Here, $\varphi_i=1/m_i^\alpha,~\alpha\in[0,1]$ is a positive known weight to account for the possible ICS scenario. We may set $\alpha=0$ by convention, which leads overweighting the large cluster sizes. When cluster sizes are correlated with the failure time, we can set the inverse of the cluster sizes, e.g., $\varphi_i = 1/m_i$, to reduce the influence of larger cluster sizes, and thus enhance statistical efficiency \citep{wang08}.

Let $\hat\bbeta^\dagger$ be the IS-BAR estimator under the multivariate PIC setting. Similar to Theorem \ref{thm1}, the asymptotic variance of $\hat\bbeta^\dagger$ can be obtained by the sandwich formula, i.e., $\var(\hat\bbeta_1^\dagger) = n^{-1}A_{n1}^\dagger(\hat\bbeta_1^\dagger)^{-1}
\Gamma_{n1}^\dagger(\hat\bbeta_1^\dagger)A_{n1}^\dagger(\hat\bbeta_1^\dagger)^{-1}$,
where
\begin{align*}
    A_n^{\dagger}(\bbeta) = \frac1{n^2}\sum_{i=1}^{n}\sum_{k=1}^{m_i}\sum_{j=1}^{n}\sum_{l=1}^{m_l}\varphi_i\varphi_{j}\eta_{2ik}\eta_{1jl}\phi\left(\frac{v_{ik}(\bbeta) - u_{jl}(\bbeta)}{r_{ikjl}}\right)\frac{(\xx_{ik} - \xx_{jl})(\xx_{ik} - \xx_{jl})^{T}}{r_{ikjl}},
\end{align*}
$\Gamma_{n}^{\dagger}(\bbeta) = \frac{1}{n}\sum_{i=1}^n\sum_{k=1}^{m_i}\{\xi_{ik}^{\dagger}(\bbeta)\}^{\otimes 2}$, where the term $\xi_{ik}^{\dagger}(\bbeta)$ can be consistently estimated by
\begin{align*}
    \xi_{ik}^{\dagger}(\bbeta) = \frac1{n}\sum_{j=1}^{n}\varphi_j\sum_{l=1}^{m_j}\biggr[&\eta_{2ik}(\xx_{ik}- \xx_{jl})I\{v_{ik}(\bbeta) < u_{jl}(\bbeta)\} \\
    &- \eta_{1jl}\left(\frac{\sum_{r=1}^n\varphi_r\sum_{o=1}^{m_r}(\xx_{ik} - \xx_{ro})I\{v_{ro}(\bbeta) > u_{jl}(\bbeta)\}}{\sum_{m=1}^n\varphi_m\sum_{t=1}^{m_m}I\{v_{mt}(\bbeta) > u_{jl}(\bbeta)\}}\right)I\{v_{ik}(\bbeta) \ge u_{jl}(\bbeta)\}\biggr],
\end{align*}
reflecting the ICS scenario. 


\section{Numerical experiments}
\label{sec3}

In this section, we conduct extensive numerical experiments using synthetic data to demonstrate the finite sample performance of the proposed IS-BAR estimator, compared to other penalties, such as the SCAD (IS-SCAD), the adaptive LASSO (IS-ALASSO) and the LASSO (IS-LASSO). Specifically, we consider three different scenarios: (a) independent covariates, (b) correlated covariates, and (c) clustered partly interval-censoring. Each simulation scenario is replicated 500 times, under sample sizes of $n=150$ and $300$.

\begin{table}[H]
\caption{Simulation results comparing the performance of the variable selection among the IS estimator, and its penalized variants (IS-BAR, IS-SCAD, IS-ALASSO and IS-LASSO) for Scenario (a) under right-censoring, with exponential and the standard normal AFT errors, and varying $n$. The Table entries are FP (average number of false non-zeros), FN (average number of false zeros), TM (probability of detecting the true model), and MSE (mean squared error). The best method for each case is highlighted in bold, and hyphens indicate zero values.. }\label{tab:moder}\vspace{-0.3in} 
\begin{center}
\renewcommand{\arraystretch}{0.7}
\begin{adjustbox}{max width=\textwidth}
\begin{tabular} { c c c c c c c c c c c c c c c c c c c c c}
\hline 
 ~& ~& ~& ~& ~& ~& \multicolumn{7}{c}{\textrm{20\% Censored}} & ~ & \multicolumn{7}{c}{\textrm{40\% Censored}} \\
\cline{7-13}\cline{15-21} 
Error &~ & n &~& Method &~ & FP &~ & FN &~& TM  &~&  MSE & ~& FP &~& FN &~& TM &~& MSE \\ 
\hline
Exp(1) &~& 150 &~& IS &~ & 7.000 &~& - &~&  0.000 &~& 0.139  & ~& 7.000 &~& - &~& 0.000 &~& 0.209  \\
~& ~& &~&   IS-BAR &~& \textbf{-} &~& \textbf{-} &~& \textbf{1.000} &~& \textbf{0.051} & ~& \textbf{-} &~& \textbf{0.015} &~& \textbf{0.990} &~& \textbf{0.094} \\
~& ~& &~&   IS-SCAD &~& 0.075 &~& - &~& 0.925 & ~& 0.058 &~& 0.175 &~& 0.055 &~& 0.800 &~& 0.222 \\
~& ~& &~&  IS-ALASSO &~& \textbf{-} &~& - &~& 1.000 & ~& 0.069 &~& \textbf{-} &~& \textbf{0.015} &~& 0.985 &~&  0.136 \\
~& ~& &~& IS-LASSO &~& 0.205 &~& 0.070 &~& 0.770 & ~& 0.713 &~&  0.125 &~& 0.935 &~& 0.320 &~& 1.890 \\
~& ~& 300 &~& IS &~& 7.000 &~& - &~& 0.000 & ~& 0.066&~& 7.000 &~& - &~& 0.000 &~& 0.101 \\
~& ~& &~&   IS-BAR &~& 0.005 &~& - &~& 0.995 & ~& \textbf{0.023} &~& \textbf{-} &~& - &~& \textbf{1.000} &~& \textbf{0.036} \\
~& ~& &~& IS-SCAD &~& 0.035 &~& - &~& 0.970 &~& 0.023 &~& 0.040 &~& - &~& 0.960 &~& 0.040 \\
~& ~& &~& IS-ALASSO &~& \textbf{-} &~& \textbf{-} &~& \textbf{1.000} &~& 0.030 &~& \textbf{-} &~& - &~& \textbf{1.000} &~& 0.049\\
~& ~& &~& IS-LASSO &~& 0.130 &~& - &~& 0.870 &~& 0.301 &~& 0.145 &~& - &~& 0.855 &~& 0.568 \\
$N(0,1)$ &~& 150 &~& IS &~& 7.000 &~& - &~& 0.000 &~& 0.091 &~& 7.000 &~& - &~& 0.000 &~& 0.119 \\ 
~& ~& &~&   IS-BAR &~& \textbf{-} &~& - &~& \textbf{1.000} & ~& \textbf{0.036} &~& \textbf{-} &~& - &~& \textbf{1.000} &~& \textbf{0.055} \\
~& ~& &~& IS-SCAD &~& 0.025 &~& - &~& 0.975 &~& 0.037 &~& 0.055 &~& 0.020 &~& 0.935 &~& 0.092 \\
~& ~& &~& IS-ALASSO &~& \textbf{-} &~& - &~& \textbf{1.000} &~& 0.049 &~& \textbf{-} &~& 0.005 &~& 0.995 &~& 0.085 \\
~& ~& &~& IS-LASSO &~& 0.175 &~& 0.010 &~& 0.830 &~& 0.503 &~& 0.135 &~& 0.600 &~& 0.470 &~& 1.487 \\
~& ~& 300&~&  IS &~& 7.000 &~& - &~& 0.000 &~& 0.101 &~& 7.000 &~& - &~& 0.000 &~& 0.054 \\ 
~& ~& &~&   IS-BAR &~& \textbf{-} &~& - &~& \textbf{1.000} &~&  \textbf{0.015} & ~& \textbf{-} &~& - &~& \textbf{1.000} &~& \textbf{0.020} \\
~& ~& &~& IS-SCAD &~& 0.015 &~& - &~& 0.985 &~& \textbf{0.015} &~& \textbf{-} &~& - &~& \textbf{1.000} &~& \textbf{0.020} \\ 
~& ~& &~& IS-ALASSO &~& \textbf{-} &~& - &~& 1.000 &~& 0.020 &~& \textbf{-} &~& - &~& \textbf{1.000} &~& 0.030 \\
~& ~& &~& IS-LASSO &~& 0.080 &~& - &~& 0.920 &~& 0.222 &~& 0.065 &~& - &~& 0.935 &~& 0.412\\
\hline
\end{tabular}
\end{adjustbox}
\end{center}
\end{table}

\subsection*{Scenario (a): Independent covariates}
Here, we generate data from the AFT model  $\log T_i = \xx_i^T\bbeta + \varepsilon_i,~i=1,\ldots,n$, where $\xx_i$ follows the $p$-dimensional multivariate standard normal distribution with the identity covariance matrix, and $\varepsilon_i$ follows either the standard normal distribution, or the unit exponential distribution. We generate the censoring distribution from the exponential distribution with rate parameter $c_0>0$ to achieve the target censoring proportions (20\% or 40\%). Table \ref{tab:moder} summarizes the simulation results with $p=10$ and $\bbeta_0 = (0,0,0,0,1,0,1,1,0,0)^T$. To evaluate the variable selection performance, we report the average number of false non-zeros (FP), the average number of false zeros (FN), the probability of detecting the true model (TM) and the mean squared error (MSE). The unpenalized IS estimator is also reported as a reference. Under the 20\% censoring proportion, most penalization methods provide satisfactory performance except for the LASSO penalty. However, both the IS-SCAD and the IS-ALASSO underperform, with all the operating characteristics seriously affected by the censoring rate. In terms of the MSE, the IS-BAR estimator appears to be about 20 times more efficient than the others under the 40\% censoring rate.

\begin{table}[H]
\caption{Simulation results comparing the finite-sample performance of the informative coefficients $(\beta_5,\beta_7,\beta_8)$ from the competing methods: IC without penalization (Induced-smoothing) and IC with BAR penalization (IS-BAR), for data generated under Scenario (a). The table entries are the bias (Bias), the empirical standard errors (ESE), the estimated standard errors (ASE), and the 95\% coverage probabilities (CP).}
\label{tab:moder-coef}
\begin{center}
\renewcommand{\arraystretch}{0.7}
\begin{tabular}{ c c c c c c c c c c c c c c c }
\hline
 ~& ~& ~& ~& \multicolumn{5}{c}{\textrm{IS}} &~& \multicolumn{5}{c}{\textrm{IS-BAR}} \\
\cline{5-9}\cline{11-15} 
Censoring rate &~& Metrics &~& $\beta_5$ &~& $\beta_7$ &~& $\beta_8$ &~& $\beta_5$ &~&
$\beta_7$ &~& $\beta_8$\\
\hline
20\% &~& Bias &~& -0.007 &~& -0.010 &~& -0.006 &~& 0.002 &~& 0.004 &~& 0.004  \\
~&~& ESE &~& 0.085 &~& 0.085 &~& 0.081 &~&  0.089 &~& 0.088 &~& 0.084 \\
~&~& ASE &~& 0.084 &~& 0.084 &~& 0.084 &~&  0.080 &~& 0.080 &~& 0.080 \\
~&~& CP &~& 0.941 &~& 0.940 &~& 0.959 &~&  0.917 &~& 0.925 &~& 0.945 \\
\\
40\% &~& Bias &~& -0.017 &~& -0.020  &~& -0.016 &~& 0.003 &~& 0.001 &~& 0.005\\
~&~& ESE &~& 0.104 &~& 0.103 &~& 0.099 &~& 0.109 &~& 0.107 &~& 0.102  \\
~&~& ASE &~& 0.106 &~& 0.106 &~& 0.106 &~& 0.100 &~& 0.100 &~& 0.100  \\
~&~& CP &~& 0.952 &~& 0.950 &~& 0.968 &~& 0.928 &~& 0.926 &~& 0.938 \\
\hline
\end{tabular}
\end{center}
\end{table}

Table \ref{tab:moder-coef} reports the performance of the IS-BAR estimator compared to the unpenalized IS estimator for the informative coefficients $(\beta_5,\beta_7,\beta_8)$, via metrics, such as the bias (Bias), the empirical standard errors (ESE), estimated standard errors (ASE) averaged over the $500$ iterations, and the coverage probabilities (CP) at the 95\% nominal confidence level. The IS-BAR estimator performs very similar to the unpenalized IS estimator. In all cases, the biases are negligible, and both estimators yield similar standard errors, with the CPs achieving the target confidence level. These results support the finite sample characteristics, as derived from Theorem \ref{thm1}-(b).

\subsection*{Scenario (b): Correlated covariates}

Here, to reflect real-world situations, we simulate cases when the covariates are correlated with each other. We retain the model in Scenario (a), while the correlation of the $l$th and $m$th components of the covariates is $r^{|l-m|}$ ($l,m=1,\ldots,10)$ with $r ~(\mbox{correlation}) =0.3$, or $0.6$. We report the simulation results in Table \ref{tab:corr}. When the covariates are correlated to each other, the IS-BAR estimator yields more satisfactory results than the other approaches.
Even though IS-SCAD and IS-ALASSO still produce small FP values, the FN and the MSE are substantially increased, compared to the independent covariates scenario. This is partly because the IS-BAR enjoys the grouping effect, as described in Theorem \ref{thm2}, which does not hold under the other penalization frameworks.

\begin{table}[H]
\caption{Simulation results comparing the variable selection performances corresponding to the competing methods with data generated under Scenario (b), for various choices of $n$ (sample size), AFT errors, and correlation ($r$). Hyphens indicate zero values.} \vspace{-0.2in} 
\label{tab:corr}
\begin{center}
\renewcommand{\arraystretch}{0.75}
\begin{adjustbox}{max width=\textwidth}
\begin{tabular} {c c c c c c c c c c c c c c c c c c c c c c c}
\hline 
 ~& ~& ~& ~& ~& ~& ~& ~& \multicolumn{7}{c}{\textrm{20\% Censored}} & ~ & \multicolumn{7}{c}{\textrm{40\% Censored}} \\
\cline{9-15}\cline{17-23} 
Correlation ($r$) &~ & AFT Error &~& n &~ & Method &~ & FP &~& FN & ~& TM &~&  MSE &~& FP &~& FN &~& TM &~& MSE \\ 
\hline
30\% &~&Exp(1) &~& 150 &~& IS-BAR &~& \textbf{-} &~& - &~& \textbf{1.000} &~& \textbf{0.062} & ~& 0.015 &~& 0.005 &~& \textbf{0.980} &~& \textbf{0.109} \\
~& ~& &~&  &~&  IS-SCAD &~& 0.070 &~& - &~& 0.930 &~& 0.061 & ~& 0.325 &~& \textbf{-} &~& 0.715 &~& 0.136 \\
~& ~& &~&  &~& IS-ALASSO &~& - &~& - &~& \textbf{1.000} &~& 0.072 & ~& \textbf{0.005} &~& 0.030 &~& 0.965 &~& 0.193 \\
~& ~& &~&  &~& IS-LASSO &~& 0.360 &~& 0.045 &~& 0.680 &~& 0.659 & ~& 0.035 &~& 2.445 &~& 0.752 &~& 2.726 \\
~& ~& &~& 300 &~& IS-BAR &~& \textbf{-} &~& - & ~& \textbf{1.000} &~& \textbf{0.026} &~& \textbf{-} &~& \textbf{-} &~& \textbf{1.000} &~& \textbf{0.043} \\
~& ~& &~&  &~& IS-SCAD &~& 0.030 &~& - & ~& 0.970 &~&0.028 &~& 0.045 &~& \textbf{-} &~&  0.955 &~& 0.047 \\
~& ~& &~&  &~& IS-ALASSO &~& \textbf{-} &~& - &~& \textbf{1.000} &~& 0.027 &~& \textbf{-} &~& \textbf{-} &~& \textbf{1.000} &~& 0.055 \\
~& ~& &~&  &~& IS-LASSO &~& 0.325 &~& - &~& 0.715 &~& 0.403 &~& 0.130 &~& 0.990 &~& 0.888 &~& 1.895 \\
~& ~& $N(0,1)$ &~& 150 &~& IS-BAR &~& \textbf{-} &~& \textbf{-} &~& \textbf{1.000} &~& \textbf{0.038} & ~& \textbf{-} &~& 0.010 &~& \textbf{0.999} &~& 0.070 \\
 &~& &~& &~& IS-SCAD &~& 0.020 &~& \textbf{-} &~& 0.980 &~& 0.039 &~& 0.060 &~& \textbf{0.005}  &~& 0.935 &~& \textbf{0.068} \\
~& ~& &~&  &~& IS-ALASSO &~& \textbf{-} &~& \textbf{-} & ~& \textbf{1.000} &~& 0.047 &~& \textbf{-} &~& \textbf{0.005} &~& 0.995 &~& 0.075 \\
~& ~& &~& &~& IS-LASSO &~& 0.325 &~& 0.035 &~& 0.710 &~& 0.649 &~& 0.340 &~& 0.530 &~&  0.450 &~& 1.282 \\ 
~& ~& &~& 300 &~& IS-BAR &~& \textbf{-} &~& \textbf{-} &~& \textbf{1.000} &~& \textbf{0.016} &~& \textbf{-} &~& \textbf{-} &~& \textbf{1.000} &~& 0.022 \\
~& ~& &~& &~& IS-SCAD &~&  0.005 &~& \textbf{-} &~& 0.995 &~& \textbf{0.016} &~& \textbf{-} &~& \textbf{-} &~& \textbf{1.000} &~& \textbf{0.019} \\
~& ~& &~& &~& IS-ALASSO &~& \textbf{-} &~& \textbf{-} &~& \textbf{1.000} &~& 0.020 &~& \textbf{-} &~& 0.645 &~& 0.936 &~& 1.195 \\ 
~& ~& &~&  &~& IS-LASSO &~& 0.140 &~& 0.010 &~& 0.985 &~& 0.302 & ~& 0.150 &~& 0.605 &~& 0.925 &~& 1.435 \\
\hline
60\% &~&Exp(1) &~& 150 &~& IS-BAR &~ & \textbf{-} &~& 0.025 &~& \textbf{0.998} &~& 0.109  & ~& 0.020 &~& 0.185 &~& 0.855 &~& 0.295 \\
~& ~& &~&  &~& IS-SCAD &~& 0.105 &~& 0.020 &~& 0.905 &~& 0.120 & ~& 0.280 &~& 0.265 &~& 0.685 &~& 0.562 \\
~& ~& &~&  &~&IS-ALASSO &~& \textbf{-} &~& \textbf{0.010} &~& 0.990 &~& \textbf{0.103} & ~& \textbf{0.005} &~& \textbf{0.070} &~& \textbf{0.925} &~& \textbf{0.284} \\
~& ~& &~&  &~& IS-LASSO &~& 0.445 &~& 1.105 &~& 0.845 &~& 1.829 & ~& 0.725 &~& 1.070 &~& 0.821 &~& 1.831 \\
~& ~& &~& 300 &~& IS-BAR &~& \textbf{-} &~& \textbf{-} &~& \textbf{1.000} &~& \textbf{0.038} & ~& \textbf{-} &~& \textbf{-} &~& \textbf{1.000} &~& \textbf{0.058} \\
~& ~& &~&  &~& IS-SCAD &~& 0.025 &~& \textbf{-} &~& 0.975 &~& 0.040 & ~& 0.040 &~& 0.015 &~& 0.965 &~& 0.086 \\
~& ~& &~&  &~& IS-ALASSO &~& \textbf{-} &~& \textbf{-} &~& \textbf{1.000} &~& 0.046 &~& \textbf{-} &~& \textbf{-} &~& \textbf{1.000} &~& 0.073 \\
~& ~& &~&  &~& IS-LASSO &~& 0.770 &~& 0.420 &~& 0.881 &~& 0.876 &~& 0.670 &~& 0.975 &~& 0.836 &~& 1.782 \\
~& ~& $N(0,1)$ &~& 150 &~& IS-BAR &~& \textbf{-} &~& 0.015 &~& \textbf{0.999} &~& 0.073 &~& \textbf{-} &~& 0.045 &~& 0.955 &~& \textbf{0.116} \\
 &~& &~& &~& IS-SCAD &~& 0.020 &~& \textbf{0.005} &~& 0.975 &~& \textbf{0.068} &~& 0.070 &~& 0.065 &~&  0.910 &~& 0.179\\ 
~& ~& &~&  &~& IS-ALASSO &~& \textbf{-} &~& \textbf{0.005} &~& 0.995 &~& 0.075 & ~& 0.005 &~& \textbf{0.020}  &~& \textbf{0.975} &~& 0.124 \\
~& ~& &~& &~& IS-LASSO &~& 0.820 &~& 0.405 &~& 0.878 &~& 0.870 &~& 0.685 &~& 1.040 &~& 0.828 &~& 1.740 \\ 
~& ~& &~& 300 &~& IS-BAR &~& \textbf{-} &~& \textbf{-} &~& \textbf{1.000} &~& 0.024 &~& \textbf{-} &~& \textbf{-} &~& \textbf{1.000} &~& 0.031 \\
~& ~& &~& &~& IS-SCAD &~& \textbf{-} &~& \textbf{-} &~& \textbf{1.000} &~& \textbf{0.022} &~& \textbf{-} &~& \textbf{-}  &~& \textbf{1.000} &~& \textbf{0.029} \\
~& ~& &~& &~& IS-ALASSO &~& \textbf{-} &~& \textbf{-} &~& \textbf{1.000} &~& 0.029 &~& \textbf{-} &~& \textbf{-} &~& \textbf{1.000} &~& 0.038 \\ 
~& ~& &~&  &~& IS-LASSO &~& 0.750 &~& 0.905 &~& 0.205 &~& 0.533 & ~& 0.810 &~& 0.865 &~& 0.833 &~& 1.626 \\
\hline
\end{tabular}
\end{adjustbox}
\end{center}
\end{table}

We further conduct simulation studies under a relatively high-dimensional covariate scenario. Here, we set the true coefficients as $\bbeta_0 = (1,1,0,0,1,1,1,\boldsymbol{0}_{p-7})^T$, where $\boldsymbol{0}_{p-7}$ denotes a $(p-7)$-dimensional all-zeros vector. Simulation results with $p=50$ are reported in Table \ref{tab:high-p50}. We observe that as the covariate dimension increases, the IS-BAR estimator still yields favorable selection performance across the various data generating processes.

\begin{table}[H]
\caption{Simulation results comparing the variable selection performances corresponding to the competing methods with data generated under Scenario (b) with high-dimensional covariates, for various choices of $n$ (sample size), AFT errors, and correlation ($r$). Hyphens indicate zero values.}
\label{tab:high-p50}
\begin{center}
\renewcommand{\arraystretch}{0.75}
\begin{adjustbox}{max width=\textwidth}
\begin{tabular} {c c c c c c c c c c c c c c c c c c c c c c c c c}
\hline 
~& ~& ~& ~& ~& ~& ~& ~& \multicolumn{7}{c}{\textrm{20\% Censored}} & ~ & \multicolumn{7}{c}{\textrm{40\% Censored}} \\
\cline{9-15}\cline{17-23} 
Corr &~ & Error &~& n &~ & Method &~ & FP &~& FN &~& TM & ~& MSE &~& FP &~& FN &~& TM &~& MSE \\ 
\hline
30\% &~&Exp(1) &~& 150 &~& IS-BAR &~ & \textbf{-} &~& \textbf{-} &~& \textbf{1.000} &~&\textbf{0.016} &~& 0.020  &~& \textbf{0.010} &~& \textbf{0.970} &~& \textbf{0.027}\\
 &~&  &~& &~& IS-SCAD &~& 0.025 &~& 0.005 &~& 0.975 &~& 0.051 & ~& 0.055 &~& 0.070 &~& 0.890 &~& 0.130 \\
 &~&  &~& &~& IS-ALASSO &~& \textbf{-} &~& \textbf{-} &~& \textbf{1.000} &~& 0.030 & ~& \textbf{-} &~& 0.035 &~& 0.965 &~& 0.065 \\
 &~&  &~& &~& IS-LASSO &~& 0.045 &~& - &~& 0.955 &~& 0.112 & ~& 0.020 &~& 0.390 &~& 0.695 &~& 0.306 \\
 &~& &~& 300 &~& IS-BAR &~& \textbf{-} &~& - & ~& \textbf{1.000} &~& \textbf{0.007} &~& \textbf{-} &~& - &~& \textbf{1.000} &~& \textbf{0.010} \\
 &~&  &~& &~& IS-SCAD &~& \textbf{-} &~& - & ~& \textbf{1.000} &~& 0.009 &~& 0.015 &~& - &~& 0.985 &~& 0.016 \\
 &~&  &~& &~& IS-ALASSO &~& \textbf{-} &~& - &~& \textbf{1.000} &~& 0.012 &~& \textbf{-} &~& - &~& \textbf{1.000} &~& 0.020 \\
 &~&  &~& &~& IS-LASSO &~& 0.015 &~& - &~& 0.985 &~& 0.053 &~& 0.005 &~& - &~& 0.995 &~& 0.090 \\
 &~& $N(0,1)$ &~& 150 &~& IS-BAR &~& \textbf{-} &~& \textbf{-} &~& \textbf{1.000} &~& \textbf{0.013} &~& \textbf{-}  &~& \textbf{-} &~& \textbf{1.000} &~&\textbf{0.016} \\
 &~& &~& &~& IS-SCAD &~& 0.025 &~& \textbf{-} &~& 0.975 &~& 0.025 &~& 0.030 &~& 0.010 &~& 0.960 &~& 0.042 \\ 
 &~&  &~& &~& IS-ALASSO &~& \textbf{-} &~& \textbf{-} & ~& \textbf{1.000} &~& 0.025 &~& \textbf{-} &~& \textbf{-} &~& \textbf{1.000} &~& 0.036 \\
 &~& &~& &~& IS-LASSO &~& 0.030 &~& 0.010 &~& 0.960 &~& 0.110 &~& 0.030 &~& 0.170 &~& 0.825 &~& 0.213 \\ 
 &~& &~& 300 &~& IS-BAR &~& - &~& - &~& 1.000 &~& \textbf{0.005} &~& - &~& - &~& 1.000 &~& \textbf{0.006} \\
 &~& &~& &~& IS-SCAD &~& - &~& - &~& 1.000 &~& 0.006 &~& - &~& - &~& 1.000 &~& 0.007 \\
 &~& &~& &~& IS-ALASSO &~& - &~& - &~& 1.000 &~& 0.010 &~& - &~& - &~& 1.000 &~& 0.013 \\ 
 &~&  &~& &~& IS-LASSO &~& - &~& - &~& 1.000 &~& 0.041 & ~& - &~& - &~& 1.000 &~& 0.056 \\
 \hline
60\% &~&Exp(1) &~& 150 &~& IS-BAR &~ & \textbf{-} &~& 0.065 &~& 0.940 &~& \textbf{0.042}  & ~& 0.050 &~& \textbf{0.115} &~& \textbf{0.860} &~& \textbf{0.068} \\
&~&  &~& &~& IS-SCAD &~& 0.125 &~& 0.165 &~& 0.755 &~& 0.193 & ~& 0.105 &~& 0.365&~& 0.625 &~& 0.302 \\
&~&  &~& &~& IS-ALASSO &~& \textbf{-} &~& 0.020 &~& \textbf{0.980} &~& 0.054 & ~& \textbf{0.030} &~& 0.160 &~&  0.820 &~& 0.114 \\
&~&  &~& &~& IS-LASSO &~& 0.195 &~& \textbf{-} &~& 0.825 &~& 0.084 & ~& 0.125 &~& 0.120 &~& 0.800 &~& 0.167 \\
&~& &~& 300 &~& IS-BAR &~& \textbf{-} &~& \textbf{-} &~& \textbf{1.000} &~& \textbf{0.013} & ~& \textbf{-} &~& 0.005 &~& \textbf{0.995} &~& \textbf{0.019} \\
&~&  &~& &~& IS-SCAD &~& 0.020 &~& 0.005 &~& 0.975 & ~& 0.022 &~& 0.010 &~& 0.025 &~&  0.965 &~& 0.061 \\
&~&  &~& &~& IS-ALASSO &~& 0.005 &~& \textbf{-} &~& \textbf{0.995} &~& 0.017 &~& 0.010 &~& \textbf{-} &~& 0.990 &~& 0.027 \\
&~& &~&  &~& IS-LASSO &~& 0.140 &~& \textbf{-} &~&  0.875 &~& 0.040 &~& 0.115 &~& \textbf{-} &~& 0.895 &~& 0.067 \\
&~& $N(0,1)$ &~& 150 &~& IS-BAR &~& \textbf{-} &~& \textbf{0.005} &~& 0.995 &~& \textbf{0.019} &~& \textbf{-} &~& \textbf{0.030} &~& 0.820 &~& \textbf{0.032} \\
&~& &~& &~& IS-SCAD &~& 0.040 &~& 0.055 &~& 0.915 &~& 0.079 &~& 0.060 &~& 0.185 &~& 0.805 &~& 0.183 \\ 
&~&  &~& &~& IS-ALASSO &~& \textbf{-} &~& \textbf{0.005} &~& \textbf{0.995} &~& 0.029 & ~& 0.005 &~& 0.055 &~& \textbf{0.945} &~& 0.060 \\
&~& &~& &~& IS-LASSO &~& 0.210 &~& \textbf{-} &~& 0.820 &~& 0.058 &~& 0.135 &~& 0.015 &~& 0.855 &~& 0.119 \\ 
&~& &~& 300 &~& IS-BAR &~& \textbf{-} &~& - &~& \textbf{1.000} &~& \textbf{0.007} &~& \textbf{-} &~& - &~& \textbf{1.000} &~& \textbf{0.010} \\
&~& &~& &~& IS-SCAD &~& 0.005 &~& - &~& 0.995 &~& 0.009 &~& \textbf{-} &~& - &~& \textbf{1.000} &~& 0.017 \\
&~& &~& &~& IS-ALASSO &~& \textbf{-} &~& - &~& 1.000 &~& 0.011 &~& \textbf{-} &~& - &~& \textbf{1.000} &~& 0.017 \\ 
&~&  &~& &~& IS-LASSO &~& 0.130 &~& - &~& 0.880 &~& 0.026 & ~& 0.055 &~& - &~& 0.945 &~& 0.045 \\
\hline
\end{tabular}
\end{adjustbox}
\end{center}
\end{table}

\subsection*{Scenario (c): Clustered partly interval-censoring}

Now, we extend the set-up in Scenario (a) to clustered PIC data.  For the marginal analysis, we generate the PIC data from the multivariate AFT model, $\log T_{ik} = \xx_{i}^T\bbeta ~ + ~ \nu_i\varepsilon_{ik},~i=1,\ldots,n;k=1,\ldots,m_i$, where $\xx_i$ follows an independent 10-dimensional multivariate standard normal distribution. The error term $\varepsilon_{ik}$ follows either the standard normal or the standard exponential distribution, and the cluster-specific random effect $\nu_i$ is generated from $N(0, 2^2)$. Cluster sizes are generated as $m_i = w/10+3$, where $w$ is the $w$-th percentile of $\nu_i$ satisfying $q_w \le \nu_i \le q_{w+10}$ for $w=0,10,\ldots,90,$ resulting in the number of subjects per cluster $m_i$ ranging from 3 to 12. We set the number of clusters as $n=70$, with $\sim$ 300 as the overall sample size. To generate the PIC observations, an interval-censoring indicator $\Delta_{ik}$ is generated from the Bernoulli distribution with probability 0.6, i.e., $P(\Delta_{ik}=1) = 0.6$, which yields approximately 40\% IC observations. We then generate a sequence of random examination times $\{S_l,~ l=1,\ldots,L\}$ by $S_{l+1} = S_{l} + \text{Unif}(0.001, 0.01)$, such that $0 = S_0< S_1 < S_2 < \ldots < S_L < \tau$, where $\tau$ is the maximum follow-up time (set to 100). When $\Delta_{ik}=0$, $U_{ik} = \max_l\{S_l:S_l \le T_{ik}\}$ and $V_{ik} = \min_l\{S_l : S_l \ge T_{ik}\}$ are observed. If $T_{ik} < U_{ik}$ or $T_{ik}>V_{ik}$, the observation is regarded as left-censored or right-censored at $U_{ik}$ or $V_{ik}$, respectively.

\begin{figure}[H]
    \centering
    \includegraphics[width=0.8\textwidth]{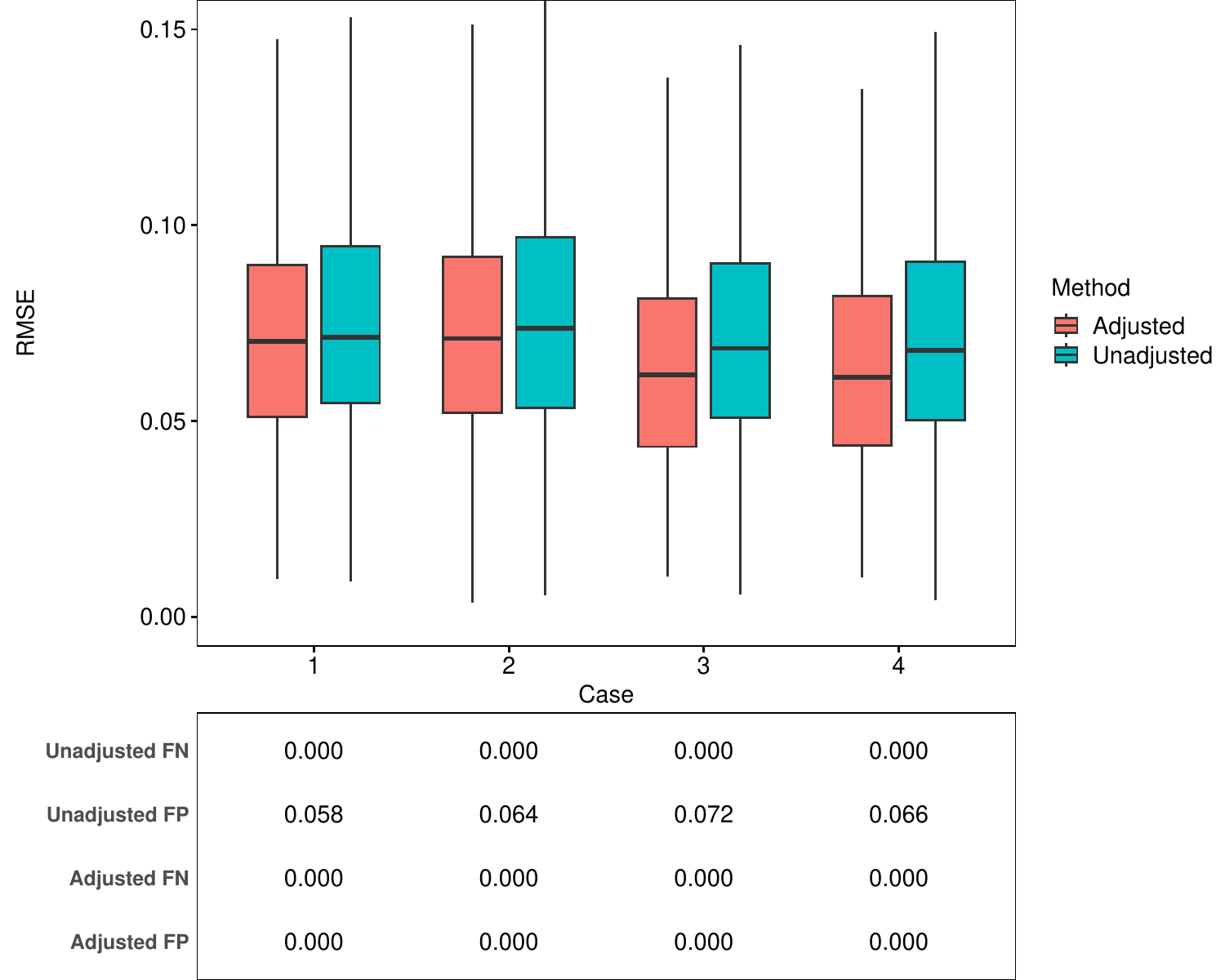}
    \caption{Estimation and variable selection performance of the ICS-adjusted and unadjusted IS-BAR methods for Scenario (c), when data are generated from the clustered partly interval-censored AFT model utilizing various combinations of the error ($\varepsilon$) and censoring rate (Cases 1--4): $\varepsilon \sim \mathrm{Exp}(1)$ and censoring rate $=20\%$ (Case 1),
$\varepsilon \sim \mathrm{Exp}(1)$ and censoring rate $=40\%$ (Case 2), 
$\varepsilon \sim N(0,1)$ and censoring rate $=20\%$ (Case 3) and 
$\varepsilon \sim N(0,1)$ and censoring rate $=40\%$ (Case 4). While the upper panel plots RMSE box-plots, the lower panel displays false-negatives (FN) and false-positives (FP).}
    \label{fig:pic-sim}
\end{figure}

We set the ICS-adjusting weight as $\varphi_i=1/m_i$, whereas $\varphi_i=1$ imposes all subjects belonging to the same cohort by imposing equal weights. To compare the performance of the ICS-adjusted method to that of the unadjusted method, we considered the combinations of two error distributions and two censoring rates. Simulation results are reported in Figure \ref{fig:pic-sim}, under
$\varepsilon \sim \mathrm{Exp}(1)$ and censoring rate $=20\%$ (Case 1),
$\varepsilon \sim \mathrm{Exp}(1)$ and censoring rate $=40\%$ (Case 2), 
$\varepsilon \sim N(0,1)$ and censoring rate $=20\%$ (Case 3) and 
$\varepsilon \sim N(0,1)$ and censoring rate $=40\%$ (Case 4).
The upper panel displays the box-plots of the root mean squared error (RMSE) for each method, while the lower panel displays the results of the selection performance using the metrics: false negatives (FN) and false positives (FP). We observe that the RMSE of the ICS-adjusted method is smaller than that of the ICS-unadjusted method across all scenarios. Furthermore, the variable selection performance of the unadjusted method worsens, compared to the adjusted method in light of larger FP values.
These results imply that the ICS-unadjusted method, which doesn't account for the latent ICS within the clustered data structure, leads to imprecise inference

\section{Applications}
\label{sec4}
In this section, we demonstrate the practical utility of the proposed method by analyzing two real-world datasets: (a) a primary biliary cirrhosis (PBC) dataset with right-censored endpoints, and (b) a metastatic colorectal cancer (mCRC) dataset with clustered partly IC endpoints. All continuous covariates were standardized in this analysis to ensure comparability across covariates. \\

\noindent \textbf{(a) Primary biliary cirrhosis data}:

We first apply our methodology to the well-known Primary Biliary Cirrhosis (PBC) dataset from the Mayo Clinic. The PBC dataset was collected from a study conducted at the Mayo Clinic between 1974 and 1984, comprising a total of 424 PBC patients, to evaluate the effect of D-penicillamine. Among them, only 312 patients had complete data, and for our analysis, we focus on 276 patients excluding missing values among these 312 patients. The censoring rate in the dataset is 59.8\%. The dataset contains 17 covariates, including seven discrete covariates (treatment, sex, ascites, hepatomegaly, spiders, edema, and stage of disease) and ten continuous covariates (age, bilirubin, cholesterol, albumin, urine copper, alkaline phosphatase, sgot, triglycerides, platelet count, and prothrombin time). 

For the PBC data analysis, we compared the unpenalized IS rank estimator with four penalized variants: IS-BAR, IS-SCAD, IS-ALASSO, and IS-LASSO. The results are summarized in Table \ref{tab:pbc}, where statistically significant covariates at the 0.05 level are marked in bold. Since all continuous covariates were standardized prior to analysis, the estimated coefficients are directly comparable in magnitude. Under the AFT model formulation, positive regression coefficients indicate longer log-survival times, whereas negative coefficients indicate shorter log-survival times, conditional on the other covariates. The unpenalized IS estimator retains all 17 covariates, and identifies several significant negative effects, including Age, Spiders, Copper, Sgot, and Prothrombin. These results suggest that, without penalization, several clinical and biochemical markers appear associated with shorter survival. However, the unpenalized estimator does not perform variable selection, and therefore may retain weak or noisy covariate effects, especially in the presence of moderate sample size and a high censoring rate.

Among the penalized approaches, IS-BAR yields the sparsest model. Specifically, IS-BAR retains only Sex, Albumin, and Platelets, with sex and Albumin statistically significant. This is consistent with the intended behavior of the BAR penalty, which approximates an $L_0$-type selection rule and aggressively removes weak signals. In contrast, IS-SCAD retains a larger set of covariates, including Treatment, Sex, Ascites, Spiders, Bilirubin, Albumin, Copper, Triglycerides, and Platelets, with Sex, Bilirubin, Albumin, and Triglycerides significant. IS-ALASSO selects several covariates, including Sex, Ascites, Spiders, Edema, Bilirubin, Copper, and Stage, but none reach statistical significance at the 0.05 level. IS-LASSO produces the least sparse penalized model, retaining most covariates, with Sex, Albumin, and Copper significant.

The Sex effect is selected across all five approaches, although its statistical significance varies. It is significant under IS-BAR, IS-SCAD, and IS-LASSO, but not under the unpenalized induced estimator, or IS-ALASSO. This suggests that Sex is a relatively robust signal in the PBC data, but its inferential strength is sensitive to the form of regularization. Albumin is another stable positive signal, selected by IS-BAR, IS-SCAD, and IS-LASSO and significant under all three methods, indicating a consistent association with longer survival. By contrast, covariates such as Ascites, Edema, Stage, and Bilirubin are selected by some methods but not others, suggesting weaker or more penalty-sensitive effects.

\begin{table}[H]
\caption{Table entries are the parameter estimates obtained from fitting the unpenalized IC (Induced) model, and its penalized variants (IS-BAR, IS-SCAD, IS-ALASSO and IS-LASSO) to the PBC data. Significant variables are marked in bold at the 0.05 significance level, and blank cells indicate coefficients shrunk to zero.} 
\label{tab:pbc}
\begin{center}
\renewcommand{\arraystretch}{0.8}
\begin{adjustbox}{max width=\textwidth}
\begin{tabular}{c rrrrrrrrrr }
\hline
Variables &~& Induced &~& IS-BAR &~& IS-SCAD &~& IS-ALASSO &~& IS-LASSO \\
\hline 
Treatment &~& --0.099 &~&  &~& 0.046 &~&  &~&  \\
Age &~& \textbf{--0.237} &~&  &~&  &~&  &~& --0.080 \\
Sex &~& 0.332 &~& \textbf{0.694} &~& \textbf{0.807} &~& 0.171  &~& \textbf{0.473} \\
Ascites &~& --0.645 &~&  &~& --0.691 &~& --0.859 &~& --0.562 \\
Hepatomegaly &~& --0.049  &~&  &~&  &~&  &~&  \\
Spiders &~& \textbf{--0.335}  &~&  &~& --0.061 &~& --0.039 &~& --0.102 \\
Edema &~& --0.606 &~&  &~&  &~& --0.933 &~& --0.150 \\
Bilirubin &~& --0.184 &~&  &~& \textbf{--0.252} &~& --0.045 &~& --0.189 \\
Cholesterol &~& --0.046 &~&  &~&  &~&  &~&  \\
Albumin &~& 0.125 &~& \textbf{0.473} &~& \textbf{0.347} &~&  &~& \textbf{0.265} \\ 
Copper &~& \textbf{--0.184} &~&  &~& --0.102 &~& --0.067  &~& \textbf{--0.144} \\ 
Alkaline &~& --0.043 &~&  &~&  &~&  &~&  \\
Sgot &~& \textbf{--0.175} &~&  &~&  &~&  &~&  \\
Triglycerides &~& 0.005 &~&  &~& \textbf{0.200} &~&  &~& 0.062 \\
Platelets &~& 0.032 &~& 0.268 &~& 0.229 &~&  &~& 0.148  \\
Prothrombin &~& \textbf{--0.159} &~&  &~&  &~&  &~& --0.062 \\ 
Stage &~& --0.241 &~&  &~&  &~& --0.071  &~& --0.061 \\
\hline
\end{tabular}
\end{adjustbox}
\end{center}
\end{table}

\begin{figure}[H]
    \centering
    \includegraphics[width=1\linewidth]{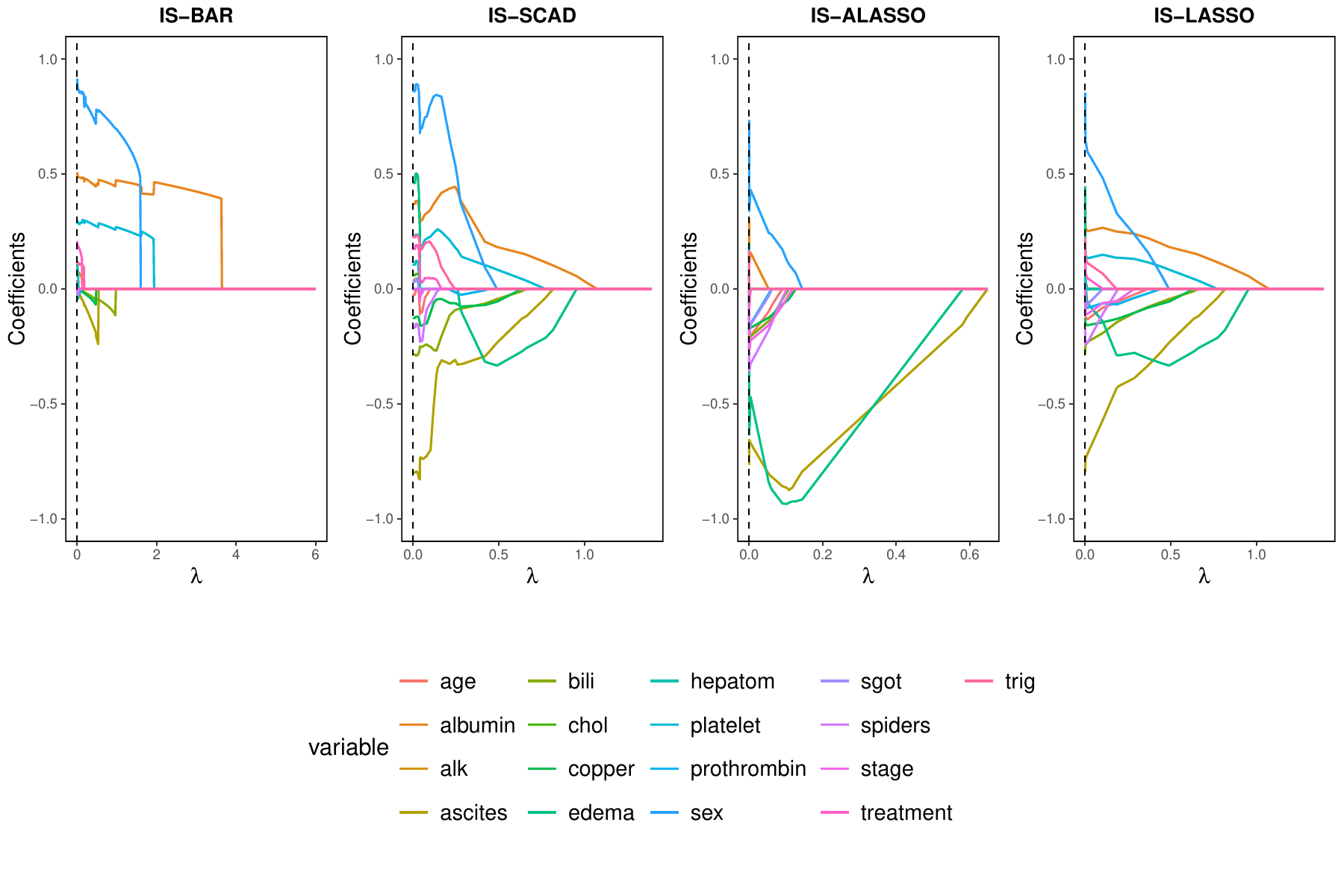}\vspace{-0.7in} 
    \caption{Solution paths of coefficients obtained from fitting the IS model (with the BAR, SCAD, ALASSO and LASSO penalties) to the PBC data.}
    \label{fig:enter-label}
\end{figure}

Figure \ref{fig:enter-label} provides additional insight through the coefficient solution paths for the four penalized estimators. The IS-BAR solution path shows a more abrupt shrinkage pattern, with most coefficients rapidly collapsing to zero, and only a small number of effects persisting across the tuning parameter range. This visually supports the sparse model reported in Table 5. By comparison, the SCAD, ALASSO, and LASSO paths show more gradual shrinkage and retain a larger number of small-to-moderate coefficients over a broader range of tuning parameters. The LASSO path, in particular, reflects its known tendency to keep many small nonzero coefficients, leading to a less parsimonious model. Overall, the combined evidence from Table \ref{tab:pbc} and Figure \ref{fig:enter-label} indicates that IS-BAR provides a compact and interpretable model for the PBC data, while preserving the most stable clinical signals, particularly Sex and Albumin.

\noindent \textbf{(b) Metastatic colorectal cancer data}:

Next, we apply our method to the clustered PIC data generated from the Amgen 20050181 study (also known as NCT00339183) --  a randomized clinical trial \citep{peeters2014} evaluating panitumumab over the standard of care (chemotherapy) in patients with metastatic colorectal cancer (mCRC), where the primary outcome is progression-free survival (PFS).  This dataset includes a wide range of covariates, including treatment allocation, tumor response, survival outcomes, adverse events and baseline characteristics. Some earlier studies chose a smaller number of covariates, e.g., age, treatment and gender \citep{pan20}.
In our analysis, we use 18 covariates (6 continuous, and 12 categorical) to investigate which of these variables are informative. The description of these variables are summarized in Table \ref{tab:mcrc_descript}. 

In this trial, tumor assessments were scheduled every eight weeks. Patients who exhibited progression (PD) at the first post-baseline assessment were treated as left-censored, whereas those whose progression was detected at later assessments were classified as IC. Patients who remained progression-free through their final assessment contributed right-censored observations, and individuals who died during the study provided exact PFS times.  After excluding patients with missing values, the total sample size was 758. Among these patients, 44 (5.80\%) had exact event times, 273 (36\%) were right-censored, 141 (18.6\%) were left-censored, and 300 (39.6\%) were IC.

\begin{table}[H]
    \caption{Description of variables in the metastatic colorectal cancer dataset}
    \begin{center}
    \renewcommand{\arraystretch}{0.8}
    \begin{adjustbox}{max width=\textwidth}
    \begin{tabular}{c l}
    \hline
    Variables & \multicolumn{1}{c}{Description} \\
    \hline
    Treatment & Actual treatment \\
    Age & Age at screening \\
    Sex & Sex \\
    Bevacizumab-M & Prior bevacizumab use (mCRC) \\
    Bevacizumab-I & Prior bevacizumab use (any setting, IVRS) \\
    Oxaliplatin-M & Prior oxaliplatin use (mCRC) \\
    Oxaliplatin-I & Prior oxaliplatin use (any setting, IVRS) \\ 
    Metastatic & Number of metastatic sites at baseline \\
    Liver & Liver-only metastasis \\
    Adjuvant & Prior adjuvant therapy use \\
    LDH & Baseline LDH level \\
    ECOG & ECOG performance status \\
    Location & Primary tumor location \\
    KRAS & KRAS mutation status \\
    EGFR & EGFR expression level \\
    BSA & Baseline body surface area \\
    Alkaline & Baseline alkaline phosphatase level \\
    Creatinine & Baseline serum creatinine level \\
    Grade & Histologic differentiation grade \\
    Adverse & Serious adverse event indicator \\
    \hline
    \end{tabular}
    \label{tab:mcrc_descript}
    \end{adjustbox}
    \end{center}
\end{table}

We fitted five penalized rank-regression models to the mCRC dataset: the ICS-unadjusted IS-BAR estimator, denoted by IS-BAR (unadj.), and four ICS-adjusted estimators, namely IS-BAR (adj.), IS-SCAD, IS-ALASSO, and IS-LASSO. The results are summarized in Table~\ref{tab:mcrc-res}, where point estimates are reported and statistically significant covariates at the 0.05 level are highlighted in bold. Since the analysis is based on the AFT formulation, positive coefficients correspond to longer progression-free survival on the log-time scale, whereas negative coefficients correspond to shorter progression-free survival, conditional on the other covariates.

Several findings emerge from Table~\ref{tab:mcrc-res}. First, the treatment effect is consistently selected by all five methods and remains statistically significant throughout, with positive estimated coefficients ranging from 0.269 (IS-LASSO) to 0.473 (ICS-adjusted IS-BAR). This provides a stable and method-insensitive signal that treatment assignment is favorably associated with PFS. Second, the BAR-based procedures remain the most parsimonious among the competing penalized estimators. Both IS-BAR (unadj.) and IS-BAR (adj.) retain only a small subset of covariates, whereas IS-ALASSO, in particular, yields a less sparse model by retaining additional variables, such as LDH and tumor location.

The comparison between the ICS-unadjusted and ICS-adjusted IS-BAR estimators is especially informative. The unadjusted IS-BAR model identifies Treatment, prior oxaliplatin use in the mCRC setting, number of metastatic sites, ECOG performance status, baseline creatinine, and histologic differentiation grade as statistically significant predictors. In contrast, after incorporating the ICS adjustment, the selected/significant profile changes: Treatment, Age, number of metastatic sites, baseline creatinine, and serious adverse event status are retained as significant, while prior oxaliplatin use, ECOG performance status, and histologic differentiation grade are no longer selected by the BAR estimator. Thus, accounting for ICS does not simply rescale the estimates; rather, it changes the inferential emphasis by attenuating some baseline disease-history and performance-status signals while highlighting Age and serious adverse event status.

\begin{table}[H]
    \caption{Results of colorectal cancer data with ICS-unadjusted IS-BAR estimator and ICS-adjusted penalized estimators. Significant variables are marked in bold at the 0.05 significance level, and blank cells indicate coefficients shrunk to zero.}
    \begin{center}
    \renewcommand{\arraystretch}{0.8}
    \begin{adjustbox}{max width=\textwidth}
    \begin{tabular}{c c c c c c}
    \hline
    & \multicolumn{4}{c}{Adjusted} & \multicolumn{1}{c}{Unadjusted} \\
    \cline{2-5} \cline{6-6}
    Variables 
    & IS-BAR (adj.)
    & IS-SCAD 
    & IS-ALASSO 
    & IS-LASSO 
    & IS-BAR (unadj.)\\
    \hline 
    Treatment       & \textbf{0.473}  & \textbf{0.359}  & \textbf{0.390}  & \textbf{0.269}  & \textbf{0.343} \\
    Age             & \textbf{0.129}  & 0.049           & 0.073           & 0.049           &  \\
    Sex             &                 &                 &                 &                 &  \\
    Bevacizumab-M   &                 &                 &                 &                 &  \\
    Bevacizumab-I   &                 &                 &                 &                 &  \\
    Oxaliplatin-M   &                 & -0.034          & --0.144          & -0.042          & \textbf{--0.043} \\
    Oxaliplatin-I   &                 &                 &                 &                 &  \\
    Metastatic      & \textbf{--0.013} & -0.047          & --0.075          & -0.050          & \textbf{--0.022} \\
    Liver           &                 &                 &                 &                 &  \\
    Adjuvant        &                 &                 &                 &                 &  \\
    LDH             & 0.158           &                 & 0.013           &                 &  \\
    ECOG            &                 & \textbf{--0.039} & \textbf{--0.071} & \textbf{--0.049} & \textbf{--0.024} \\
    Location        &                 &                 & 0.034           &                 &  \\
    KRAS            &                 &                 &                 &                 &  \\
    EGFR            &                 &                 &                 &                 &  \\
    BSA             &                 &                 &                 &                 &  \\
    Alkaline        &                 &                 &                 &                 &  \\
    Creatinine      & \textbf{0.147}  & 0.049           & 0.049           & 0.043           & \textbf{0.112} \\
    Grade           &                 &                 &                 &                 & \textbf{0.080} \\
    Adverse         & \textbf{--0.015} & \textbf{--0.389} & \textbf{--0.428} & \textbf{--0.297} &  \\
    \hline
    \end{tabular}
    \end{adjustbox}
    \end{center}
    \label{tab:mcrc-res}
\end{table}

The adjusted methods also reveal several patterns of robustness and penalty sensitivity. Treatment is the most stable positive signal, being selected and significant across all approaches. Serious adverse event status is selected only by the ICS-adjusted methods and is consistently negative and statistically significant, suggesting that once ICS is accounted for, adverse event status becomes an important marker of shorter progression-free survival. The number of metastatic sites is selected by all methods, but it is statistically significant only under the BAR-based estimators, indicating that its inferential strength depends on the penalty structure. Similarly, baseline creatinine is retained by all methods, but achieves statistical significance only under IS-BAR (unadj.) and IS-BAR (adj.). ECOG performance status shows the opposite pattern: it is significant under IS-BAR (unadj.) and under the adjusted L1-type penalties, but not under ICS-adjusted IS-BAR, suggesting that this signal is more sensitive to the stronger sparsity induced by the BAR penalty. Overall, Table~\ref{tab:mcrc-res} supports the practical value of the ICS-adjusted IS-BAR approach: it produces a compact and clinically interpretable model while emphasizing predictors that remain important after accounting for the clustered PIC structure of the data.

\section{Conclusions}
\label{sec5}

In this paper, we develop an $L_0$-type broken adaptive ridge (BAR) penalization framework for semiparametric accelerated failure time (AFT) models with right-censored and multivariate partly interval-censored data. The proposed method combines the sparsity-inducing behavior of BAR with an induced-smoothing version of the Gehan-type rank objective function. This construction avoids direct optimization of the nonsmooth rank-based objective, and eliminates the need for external bandwidth selection, since the smoothing scale is naturally determined within the induced-smoothing formulation. To further enhance computational scalability, we proposed a cyclic coordinate descent (CCD) algorithm based on a local least-squares representation of the smoothed objective. This algorithm enables fast and stable coefficient updates, while also providing a unified computational framework that can accommodate other commonly used penalties, including LASSO, adaptive LASSO, and SCAD.

The proposed IS-BAR estimator retains the main theoretical advantages of BAR-type regularization. In particular, we established the oracle property and the grouping effect, as in other BAR-based estimators \citep{Dai2018,kawa21,Leechoi24}, thereby showing that the method can consistently identify the informative covariates while producing stable estimates in the presence of highly correlated predictors. These properties are especially relevant in biomedical and clinical studies, where covariates often exhibit substantial correlation and where parsimonious, interpretable models are desirable. Through extensive numerical experiments, the proposed method demonstrated superior variable-selection accuracy and estimation efficiency compared to several competing $L_1$-based penalization approaches. The real-data analyses further illustrated its practical utility in clinically relevant survival settings. 

Several extensions remain important directions for future research. First, modern clinical datasets, particularly those arising from electronic health records, disease registries, and multi-center studies, often involve massive sample sizes, high-dimensional covariate spaces, and complex dependence structures. Extending the proposed IS-BAR framework to settings where the covariate dimension diverges with sample size would broaden its applicability to large-scale biomedical data. Here, distributed, divide-and-combine strategies \citep{wang2022multivariate} offer a promising route for scaling the method to massive datasets, where the full dataset may be partitioned into multiple sub-samples, with penalized estimators computed locally and subsequently aggregated. A key statistical challenge is how to combine both the estimated coefficients and the selected informative variables across sub-samples without sacrificing selection consistency, or full-sample efficiency. One possible approach is to incorporate a one-step linear approximation or bias-correction step \citep{hong22}, which may recover full-sample efficiency from aggregated sub-sample estimators. Further theoretical work is needed to establish the validity of such procedures under censoring, interval censoring, and ICS mechanisms.

Additional future directions include extending the proposed framework to time-dependent covariates, competing risks, recurrent events, and multistate survival outcomes. These settings arise frequently in longitudinal clinical studies, and would allow the BAR-penalized rank-regression framework to address a broader class of event-history data. It would also be valuable to develop post-selection inference procedures and stability-assessment tools for the selected variables, particularly in high-dimensional applications, where model uncertainty may be substantial. Overall, the proposed IS-BAR method provides a flexible, theoretically justified, and computationally efficient platform for sparse semiparametric survival regression, with substantial potential for further development in complex biomedical data analysis.

\section*{Acknowledgments}
T. Choi was supported by the National Research Foundation of Korea (NRF) grant funded by the Korean government (RS-2024-00340298). D. Bandyopadhyay acknowledges funding support from grants P30CA016059, 	R01DE031134 and R21DE031879 from the United States National Institutes of Health. 
S. Park was supported by the government of the Republic
of Korea (MSIT) and the National Research Foundation of Korea (RS-2024-00338876).
D. Kim was supported by the government of the Republic of Korea (MSIT) and the National Research Foundation of Korea (RS-2025-24683613).

\bibliographystyle{apalike}
\bibliography{biblist}

\end{document}